\documentclass[letterpaper, 10 pt, conference]{ieeeconf}  
\usepackage[table,x11names]{xcolor}
\usepackage{hyperref}

\IEEEoverridecommandlockouts                              

\usepackage[colorinlistoftodos,bordercolor=orange,backgroundcolor=orange!20,linecolor=orange,textsize=scriptsize]{todonotes}
\usepackage{amsmath,amssymb,amsfonts,mathtools}%
\usepackage{multirow}

\usepackage{amsthm}
\usepackage{hyperref}
\usepackage{cleveref}
\usepackage{mathabx}
\usepackage{dsfont}
\DeclareMathOperator*{\argmax}{arg\,max}

\DeclareMathOperator*{\sgn}{sgn}

\renewcommand{\geq}{\geqslant}
\renewcommand{\leq}{\leqslant}

\newcommand{\conv}{\operatorname{conv}}
\newcommand{\Trav}{\operatorname{Trav}}

\def\<#1,#2>{\langle #1,#2\rangle}
\newtheorem{theorem}{Theorem}
\newtheorem{proposition}[theorem]{Proposition}%
\newtheorem{defprop}[theorem]{Definition-Proposition}
\newtheorem{corollary}[theorem]{Corollary}%
\newtheorem{lemma}{Lemma}

\theoremstyle{definition}%
\newtheorem{example}{Example}%

\newtheorem{definition}{Definition}%
\newtheorem{assumption}{Assumption}%

\usepackage{soul}
\usepackage{xcolor}
\setstcolor{red}

\title{\LARGE \bf
Dynamics of multidimensional Simple Clock Auctions
}

\author{Jad Zeroual$^{1, 2}$, Marianne Akian$^{1}$, Aurélien Bechler$^{2}$, Matthieu Chardy$^{2}$, Stéphane Gaubert$^{1}$ 
\thanks{$^{1}$ CMAP - Inria, Ecole Polytechnique, Route de Saclay, 91128 Palaiseau, France
        }
\thanks{$^{2}$Orange Labs, 46 Av. de la République, 92320 Châtillon, France
}}

\begin{document}

\maketitle
\thispagestyle{empty}
\pagestyle{empty}

\begin{abstract}
Simple Clock Auctions (SCA) are a 
mechanism commonly used in spectrum auctions to sell lots of frequency bandwidths. 
We study such an auction with one player having access to perfect information against straightforward bidders. When the opponents' valuations satisfy the
ordinary substitutes condition, we show that it is optimal to bid on a fixed lot over time. In this setting, we consider a continuous-time version of the SCA auction in which the prices follow a differential inclusion with a piecewise-constant dynamics. We show that there exists a unique solution in the sense of Filippov. This guarantees that the continuous-time model coincides with the 
limit of the discrete-time auction when price increments tend to zero. Moreover, we show that the value function of this limit auction is piecewise linear (though possibly discontinuous). Finally, we illustrate these results by analyzing a simplified version of the multiband Australian spectrum auction of 2017.
\end{abstract}

\section{Introduction}

\subsection{Context}
Clock auctions are sequential multiround auctions where the prices of items or bundles increase over time, following a clock (e.g. the number of rounds). They were at the heart of the literature during the 2000s, as they were praised for their nature to facilitate price discovery~(\cite{milgrom_putting_2004, cramton_ascending_1998, crampton_clock}) and result in efficient and fair allocations~(\cite{milgrom_putting_2004,book_klemperer}). As a consequence, those formats gained popularity among spectrum auctions during the 4G and 5G era~(\cite{nera, cramton_spectrum_2009}). In particular, the Simple Clock Auction (SCA) format has been prolific
in the 2010s~\cite{nera}. It is renowned for its simplicity and practicality among professionals~(\cite{crampton_clock, dotecon_advice}) 
and could be interesting in future auctions when licenses for the allocated bandwidths expire. Indeed,
it allows the simultaneous auctioning of heterogeneous bandwidths,
which is one of the plausible scenarios for future auctions in France for instance~\cite{consultation_publique}.

\subsection{Contributions}
We adopt a perfect information setting in which one player knows the payoffs of all their opponents. We also assume that each opponent plays according to the straightforward bidding strategy (SB). This strategy
(described in~\cite{cramton_ascending_1998, bikhchandani_competitive_1997, ausubel_milgrom}) consists in maximizing the profit given
the current price of the auction without anticipating future prices evolution, i.e.\ as if the auction would stop at the current round. 
Provided the valuations of the opponents satisfy 
the {\em ordinary substitutes} condition, we show in \Cref{constant_optimality} that it is optimal to bid on a constant bundle of items over time. The ordinary substitutes condition
is a basic property in the theory of combinatorial auctions, it means
that increasing the price of one category of goods does not decrease the demand
for other goods (\cite{milgrom_substitute_2009,baldwin_klemperer}).
The proof 
of \Cref{constant_optimality} is based on a formulation in terms of deterministic optimal control
(\Cref{DP}).

Then, we study the dependency of the value of the auction in the price increments. To do so, 
we introduce a continuous-time version of the auction, obtained
by letting price increments tend to zero. This leads to a differential inclusion for the price trajectory.
Exploiting the ordinary substitutes condition together with tropical geometry methods~(\cite{baldwin_klemperer, tran_yu_2019,mikhalkin_decomposition_2004,Itenberg2009}),
we show that a solution of the differential inclusion exists and is unique in the sense of Filippov~\cite{filippov_differential_1988}
(\Cref{existence} and \Cref{uniqueness}).
That is done by using the sup-norm as a Lyapunov function~\cite{aubin_differential_1984}. It follows that our model is a limit of the SCA when the price increments tend to zero. 
We also prove that the value function of the continuous time problem is piecewise linear but not necessarily continuous (\Cref{th-valueissemilinear}). Moreover, this value function coincides with the limit of the value functions in discrete-time (\Cref{coro-approx}).

We exemplify our results by an application on a simplified version of the multiband Australian auction of 2017~\cite{auction_results}, with only two operators
and two categories of bandwidths. We compute the optimal payoff
of each operator, assuming their competitor plays straightforward bidding.
This allows us to compute a constant optimal strategy and to exhibit the optimal dynamics of prices for each competitor.
\subsection{Related work}
Baldwin and Klemperer have recently used tropical geometry to analyze the agent demand types in combinatorial auctions and establish conditions on the existence of equilibria (see~\cite{baldwin_klemperer}). This has been further elaborated by Tran and Yu~\cite{tran_yu_2019}. This is closely related to the discrete convexity theory developed by Danilov, Koshevoy, and Murota (\cite{danilov_discrete_2004, murota_discrete_2003,murota_discrete_2016}).  In this approach, the demand of a player as a function of the price is represented by a {\em tropical hypersurface}, a special kind of polyhedral complex
arising as a log-limit of a family of classical hypersurfaces (\cite{mikhalkin_decomposition_2004,Itenberg2009}). 
Baldwin and Klemperer considered the tropical hypersurfaces originating from ordinary substitutes valuations. These valuations play an essential role in auction theory as they guarantee the
existence of a particular notion of equilibrium in clock auctions~(\cite{milgrom_putting_2004, milgrom_substitute_2009}).
Here, we apply the tropical approach to a different problem:
the analysis of the price dynamics and optimal strategies
against straightforward bidders in a SCA.
\Cref{constant_optimality} extends to the multidimensional case~\cite[Proposition~4]{zeroual_2025} showing that a constant strategy is optimal in one-dimensional SCA. 

Classical references on differential inclusions include the books by Filippov~\cite{filippov_differential_1988} and Aubin and Cellina~\cite{aubin_differential_1984}.
They establish the existence of solutions to differential equations with discontinuous vector fields
by extending the notion of Carathéodory solutions. Uniqueness, on the other hand, is shown under restrictive assumptions, including
Lipschitz continuity 
of the vector field~\cite{filippov_differential_1988} or its monotonicity~(\cite{aubin_differential_1984, peypouquet_evolution_2010}).
However, in our framework, both assumptions fail.
A more closely related result is \cite[Theorem~2, \S 10, Ch.~2]{filippov_differential_1988}, dealing with the case of a single surface of discontinuity.
The difficulty here is the presence of {\em several} surfaces (cells of a polyhedral complex). Then, to obtain
a global uniqueness result (\Cref{uniqueness}),
we rely on geometrical properties taking stem in the auction mechanism
and the ordinary substitutes condition. This makes the auction dynamics tractable. In contrast, we recall that the dynamics of piecewise constant vector fields
is generally complex. In particular the reachability problem is undecidable~\cite{Asarin1995} and even belongs to higher classes of the arithmetic hierarchy~\cite{Bournez1999}. 

A recent survey on dynamical systems with discontinuous 
vector fields can be found in~\cite{4518905}. We also refer to the monograph on switched systems~\cite{sun_switched_2005} for general information.
Proofs of uniqueness of trajectories for specific models have appeared in~\cite{danca_class_2001} and in~\cite{machina_filippov_2011} (with an application to mathematical biology).

\subsection{Notation}

We denote by $\mathbb{R}$ the set of real numbers and by $\mathbb{Z}$ the set of integers.
The notation $A_{\geq 0}$ (resp.\ $A_{> 0}$) stands for the nonnegative (resp.\ positive) elements of set $A$. The cardinal of $A$ is denoted by $\# A$. For $k \in \mathbb{Z}_{> 0}$, we write $[k]\coloneqq \{1, \dots, k\}$. We denote by $e_i \in \mathbb{R}^M$ the vector with a $1$ in its $i^{\text{th}}$ component and zero elsewhere and by $\langle ., . \rangle$ the canonical scalar product of $\mathbb{R}^M$.
We use $\mathds{1}$ for the indicator function, e.g. $\mathds{1}_{x > 0}$ is equal to $1$ if $x > 0$ and $0$ otherwise. In the sequel, $\dot x (t)$ symbolizes the time derivative of $x$ at time $t$ and $\sgn$ the function defined on $\mathbb{R}\backslash\{0\}$ such that $\sgn(y) = \frac{y}{|y|}$. 

\section{Multidimensional clock auction}

\subsection{Auction mechanism}
We model a multi-dimensional SCA between {$N$} players by a tuple 
{$(M, m, p_{\min}, \epsilon,v_1,\ldots,v_N)$} where: 
\begin{itemize}
    \item $M$ is the number of item categories put up for auction;
    \item $m$ is a vector in $\mathbb{R}^M$
      such that $m_j = \langle m, e_j\rangle \in \mathbb{Z}_{> 0}$ is the number of items available in category $j \in [M]$;
    \item $p_{\min} \in \mathbb{R}_{\geq 0}^M$ is the initial price and $\epsilon \in \mathbb{R}_{> 0}^M$ the price increment of the auction;
    \item 
    $v_i$ for $i\in [N]$ are real-valued functions 
    on $\mathcal{M} \coloneqq \prod_{j = 1}^M\{0, \dots, m_j\}$; they are the players' \textit{valuations}. 
\end{itemize}
At each round, at price $p\in\mathbb{R}_{\geq 0}^M$, each player 
{$i\in [N]$} asks for a bundle of items $d^{(i)}
\in \mathcal{M}$. 
Then, if 
{$D\coloneqq\sum_{i\in [N]} d^{(i)}$ satisfies $D_j \leq m_j$} for all $j \in [M]$, the auction terminates and player $i$ receives {$d^{(i)}$}. Otherwise, the prices of categories in excess increase according to the increment vector $\epsilon$ 
(i.e.\ 
{$\forall j \in [M], p'_j = p_j + \epsilon_j \mathds{1}_{D_j > m_j}$}) and the auction continues for another round.
This extends the SCA described in~\cite{zeroual_2025} (case $M=1$).


\subsection{Valuations}
In the context of spectrum auctions, we interpret 
the valuations as the budget for each bundle of items.
It is standard in auction theory to model preferences through these functions (see~\cite{bikhchandani_competitive_1997, ausubel_milgrom, milgrom_putting_2004, milgrom_substitute_2009, baldwin_klemperer}). 
We assume that 
{a valuation $v$ is} non-decreasing, non-negative and such that 
{$v(0)=0$}. 
Furthermore, we assume that the payoff 
for 
{a player with valuation $v$} acquiring a bundle $k\in\mathcal{M}$ at price $p\in\mathbb{R}^M_{\geq 0}$ is 
$v(k) - \langle k, p \rangle $. 

We will also assume that the opponents are Straightforward bidders (see next section). This allows us to perfectly aggregate opponents~\cite{zeroual_2025}.
We therefore {assume that there are only $2$ players}, calling player $1$ \textit{the player} and player $2$ \textit{the opponent}. {We denote by $w$ (resp.\ $v$) the valuation of the player (resp.\ the opponent).} 


\subsection{Auction against Straightforward bidding (SB)}

We suppose the opponent plays SB~(\cite{milgrom_putting_2004,bikhchandani_competitive_1997, milgrom_substitute_2009}). Our goal is to compute an optimal strategy for the player in this case.

\begin{definition}
    The demand of the SB opponent 
    at price $p \in \mathbb{R}^M_{\geq 0}$ is $\mathcal{D}(p) = \argmax_{\delta \in \mathcal{M}}\{  v(\delta) - \langle \delta, p \rangle\}$.
    When this set is single-valued, we identify $\mathcal{D}(p)$ to 
    its sole element $\delta$. In this case, we define $\mathcal{D}_j(p) \coloneqq \langle \delta, e_j \rangle$ for all $j \in [M]$.
\end{definition}

The set $\mathcal{T} \coloneqq  \{ p \in \mathbb{R}^M_{\geq 0} : \#\mathcal{D}(p) > 1 \}$ is the set of prices where the maximum of the payoff is attained at least twice. This is known as the {\em indifference locus}~\cite{baldwin_klemperer}, for at any price $p\in \mathcal{T}$, two bundles must be equally attractive.
The set $\mathcal{T}$ coincides with the nondifferentiability locus of the piecewise-linear map $p \mapsto \max_{k\in \mathcal{M}} (v(k)-\<k,p>)$, this is known as a {\em tropical hypersurface},
a special kind of polyhedral complex that plays
a key role in tropical geometry~\cite{mikhalkin_decomposition_2004,Itenberg2009}.
An example of tropical hypersurface is shown in \Cref{fig:tropical_curve}.
\begin{figure}[!htbp]
    \centering
\includegraphics[width=0.5\linewidth]{tropical_curve_ex_1.jpg}
    \caption{An example of demand function $\mathcal{D}$, with $M = 2$ and $m_1 = m_2 = 3$, and the associated tropical hypersurface. The demand takes a constant value on the interior of each maximal cell.}
    \label{fig:tropical_curve}
\end{figure}

It 
{defines} a collection of {\em cells} (closed convex polyhedra, see~\cite[Def.\ 1]{mikhalkin_decomposition_2004}). 
Cells vary in sizes. Maximal cells have the dimension of their affine span coincide with the dimension of the ambient space (in~\Cref{fig:tropical_curve}, they are the hexagons and the semi-infinite regions like the one labeled $(0,3)$). Cells also refer to lower dimensional polyhedra such as the segments of $\mathcal{T}$ at the intersection of two neighboring hexagons in~\Cref{fig:tropical_curve}. Maximal cells form a covering of $\mathbb{R}^M_{\geq 0}$ and only one bundle is demanded in their interiors.
%

What's more for every $\delta \in \mathcal{M}$, the subset $\mathcal{D}^{-1}(\delta) \coloneqq \{ p \in \mathbb{R}^M_{\geq 0} : \mathcal{D}(p) \ni \delta \}$ is a {nonempty} polyhedron. In the following, we will assume that $v$ is strictly concave to ensure that every cell $\mathcal{D}^{-1}(\delta) $ has a non-empty interior (and so, is maximal).

The following result is straightforward.
\begin{defprop}\label{max_price}
    The price $p_{\max}$ defined by $p_{\max, j} \coloneqq  1+\max_{\substack{\delta \in \mathcal{M} \\ \delta_j \geq 1}} v(\delta)$ for all $j \in [M]$ satisfies $\forall p \in \mathbb{R}^M_{\geq 0}\backslash\mathcal{T}$, $\forall j \in [M]$,$(p_j \geq p_{\max, j} \implies \mathcal{D}_j(p) = 0)$.
\end{defprop}
\begin{definition}
  Let $p \in \mathbb{R}^M_{\geq 0}\backslash\mathcal{T}$ be the price of the auction at a round. The price in the following round is given by
  \[ T(k, p, \epsilon) = p + \begin{pmatrix}
        \epsilon_1 \mathds{1}_{k_1 + \mathcal{D}_1(p) > m_1}\\
        \vdots\\
        \epsilon_M \mathds{1}_{k_M + \mathcal{D}_M(p) > m_M}
  \end{pmatrix}
  \]
  where $k \in \mathcal{M}$ is the demand of the player at price $p$.
\end{definition}

The set $\mathcal{T}$ has zero measure;
so a sequence of prices will avoid it
if the valuation is generic {and $p_{\min}\not\in \mathcal{T}$}. Thus assuming that the prices never lie on $\mathcal{T}$ is not a heavy assumption~\cite{milgrom_substitute_2009}. 
{Hereafter, we denote by $G \coloneqq \prod_{j = 1}^M (p_{\min, j} +  \mathbb{Z}_{\geq 0}\epsilon_j)$ the grid of prices starting at $p_{\min}$ assuming it avoids $\mathcal{T}$: $G\subset \mathbb{R}^M_{\geq 0}\setminus \mathcal{T}$.}


\begin{definition}\label{sequence}
    For a sequence of demands $k = (k^n) \in \mathcal{M}^{\mathbb{Z}_{\geq 0}}$, we define the sequence $(p^n)_{n \in \mathbb{Z}_{\geq 0}} \in G^{\mathbb{Z}_{\geq 0}}$ such that $p^0 = p \in G$ and $\forall n \in \mathbb{Z}_{\geq 0}, p^{n+1} = T(k^n, p^n, \epsilon)$. We denote by $T^n(k, p, \epsilon)$ the $n^{th}$ element $p^n$ of the price sequence. 
\end{definition}

In this definition, we can observe that $(k^n)$ is the sequence of the player's bids and $(\mathcal{D}(p^n))$ is the opponent's. 
The sequence $(k^n)$ can be viewed as the \textit{strategy} of the player. Indeed, since the opponent plays SB, a player's strategy is fully characterized by the sequence of bids. 
The dynamics of prices 
is given by $(T^n(k, p_{\min}, \epsilon))_{n \in \mathbb{Z}_{\geq 0}}$.

\begin{proposition}\label{termination}
    (Auction termination) Let $k = (k^n)_{n \in \mathbb{Z}_{\geq 0}}$. There exists $N \in \mathbb{Z}_{\geq 0}$ such that $k^N_j + \mathcal{D}_j(T^N(k, p_{\min}, \epsilon)) \leq m_j$ for all $j \in [M]$. 
\end{proposition}
\begin{proof}
    Let $\forall n \in \mathbb{Z}_{\geq 0}$, $p^n = T^n(k, p_{\min}, \epsilon)$. Suppose by contradiction 
    that $\forall n \in \mathbb{Z}_{\geq 0}, \exists j \in [M], k^n_j + \mathcal{D}_j(p^n) > m_j$. There exists a function $\psi : \mathbb{Z}_{\geq 0} \rightarrow [M]$ such that for all integer $n, k^n_{\psi(n)} + \mathcal{D}_{\psi(n)}(p^n) > m_{\psi(n)}$. There exists $j \in [M]$ such that $\psi^{-1}(\{j\})$ is not finite: let $n_j \in \mathbb{Z}_{\geq 0}$ such that $ \#\{n \leq n_j : \psi(n) = j\} > \frac{p_{\max, j} - p^0_j}{\epsilon_j}$. Hence, $\forall n \geq n_j, p^n_j \geq p_{\max, j}$ giving $\forall n \geq n_j, k^n_j + \mathcal{D}_j(p^n) = k^n_j \leq m_j$ which is a contradiction. 
\end{proof}
\subsection{Optimal strategy against SB}\label{values}

If the player knows the valuation of the opponent, then they could compute their value defined as, for $k' \in \mathcal{M}$ and $p \in \mathbb{R}^M_{\geq 0}$:
\[ W(k',p) = \max_{\substack{\kappa=(k^n) \in \mathcal{M}^{\mathbb{Z}_{\geq 0}}\\ (\|k^n\|_1) \, \text{non-increasing} \\ k^0 = k'}} w(k^{N_k}) - \langle k^{N_k}, T^{N_k}(\kappa, p, \epsilon) \rangle
\]
where $N_\kappa$ is the round at which the auction stops if the player plays according to strategy $\kappa$. 
Notice that the sum of demands should not increase. This is known as the \textit{eligibility rule}. 
The value can be computed thanks to dynamic programming.

\begin{figure}[!htbp]
    \centering
    \includegraphics[width=0.5\linewidth]{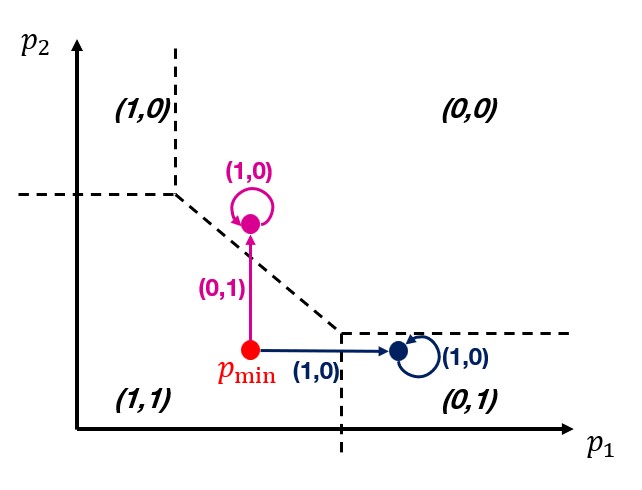}
    \caption{Illustration of a non-constant optimal strategy when $w(1,0) \gg w(k)$. The tropical hypersurface associated to the valuation of the opponent is represented (dashed lines).}
    \label{fig:constant_strat}
\end{figure}

\begin{proposition}\label{DP}(The DP algorithm~{\cite[Prop.~3.1]{bertsekas}})
  The value of the problem satisfies the following equation:
  for all $k \in \mathcal{M}$ and $p \in G$, 
  $W(k,p) = w(k) - \langle k, p \rangle$ if for all j $\in [M]$, $k_j + \mathcal{D}_j(p) \leq m_j$ and $W(k,p) = \max_{\substack{l \in \mathcal{M} \\ \|l\|_1 \leq \|k\|_1}} W(l, T(k,p, \epsilon))$ otherwise.
\end{proposition}
For $M=1$, a constant strategy (bidding a quantity of goods that is constant over time) is optimal (\cite{zeroual_2025}). This is not the case in general as shown by the following example.
\begin{example}
  Let us consider a two-dimensional {SCA} 
  Suppose that for all $k \neq (1,0)$, $w(1, 0) \gg w(k)$ so that the player must end the auction with demand $(1,0)$.
  Let $v(0,0) = 0$, $v(1, 0) = 1$, $v(0, 1) = 2$, $v(1, 1) = 5$ and $\epsilon \in \mathbb{R}^2_{> 0}$ such that $\epsilon_2 > \epsilon_1$. We set the starting price to $p_{\min} = (3 - \frac{\epsilon_1}{2}, 2 - \frac{\epsilon_2}{2})$. Bidding $(1,0)$ would result in a new price $T((1,0), p_{\min}, \epsilon) = p_{\min} + (\epsilon_1, 0) = (3 + \frac{\epsilon_1}{2}, 2 - \frac{\epsilon_2}{2})$.
  At this price,
  the opponent asks for the bundle $(0,1)$. Therefore, the auction stops
  if the player asks for $(1,0)$ again, yielding a payoff of $w(1,0) - 3 - \frac{\epsilon_1}{2}$. Meanwhile, bidding $(0,1)$ at $p_{\min}$ results in a price of $(3 - \frac{\epsilon_1}{2}, 2 + \frac{\epsilon_2}{2})$ where the opponent asks for $(0,0)$. 
  Bidding $(1,0)$ would therefore end the auction with a final payoff $w(1,0) - 3 + \frac{\epsilon_1}{2} > w(1,0) - 3 -\frac{\epsilon_1}{2}$. This is illustrated in \Cref{fig:constant_strat}.
\end{example}
\section{Optimality of a constant strategy}\label{section3}

In this section, we show that, under a classical assumption over the opponent's valuation, a constant strategy is optimal regardless of the number $M$
of item categories.

\begin{definition}\label{def_v}
  We define $V(k,p)$ as the payoff of the player obtained by bidding a constant demand $k$ throughout the auction, starting from price $p$: 
  $V(k,p) = w(k) - \langle k, p \rangle$ if $\forall j \in [M], k_j + \mathcal{D}_j(p) \leq m_j$ and $V(k, T(k,p,\epsilon))$ otherwise.   

  We define $\tilde{V}$ as the optimal payoff over the set of strategies that remain constant from round~$1$, i.e.
  $\Tilde{V}(k,p) = w(k) - \langle k, p \rangle$ if $\forall j \in [M], k_j + \mathcal{D}_j(p) \leq m_j$ and $ \max_{\substack{l \in \mathcal{M} \\ \|l\|_1 \leq \|k\|_1}} V(l,T(k,p,\epsilon))$ otherwise.
\end{definition}
A constant strategy is optimal if and only if $W = \Tilde{V}$. 
\subsection{Ordinary substitutes}

The following definition is taken from
Milgrom~(\cite{milgrom_putting_2004, baldwin_klemperer}):

\begin{definition}\label{substitute}
    Goods are {\em substitutes} for the opponent if, on $\mathbb{R}^M_{\geq 0}\backslash\mathcal{T}$, for any good category $j$, increasing the prices of other categories does not reduce the demand for category $j$: {$[\hat{p}_{i} \geq p_{i}\; \forall i\neq j,\;\text{and}\; \hat{p}_j = p_j]\implies [\mathcal{D}_j(\hat{p}) \geq \mathcal{D}_j(p)]$.}
\end{definition}

This property of the valuation conveys the intuitive idea that a price raise in one category would induce a raise in demands on the other categories.

\begin{proposition}\label{increasing_T}
Let $p, \hat{p} \in G$ such that $\hat{p} \geq p$. If goods are substitutes for the opponent, then for all $k \in \mathcal{M}, T(k,\hat{p}, \epsilon) \geq T(k,p,\epsilon)$.
\end{proposition}
\begin{proof}
    Suppose goods are substitutes for the opponent. Let $j \in [M]$. 
    If $\hat{p}_j > p_j$ then $\hat{p}_j \geq p_j + \epsilon_j$, because
    $\hat{p}$ and $p$ belong to the grid $G$. Therefore, $T_j(k, \hat{p}, \epsilon) - T_j(k,p,\epsilon) \geq \epsilon_j(1 + \mathds{1}_{k_j + \mathcal{D}_j(\hat{p}) > m_j}-\mathds{1}_{k_j + \mathcal{D}_j(p) > m_j})\geq 0$.
    Otherwise, \Cref{substitute} implies $\mathcal{D}_j(\hat{p}) \geq \mathcal{D}_j(p)$. Therefore, $\mathds{1}_{k_j + \mathcal{D}_j(\hat{p}) > m_j}$ is greater or equal to $\mathds{1}_{k_j + \mathcal{D}_j(p) > m_j}$, thus $T_j(k, \hat{p}, \epsilon) \geq T_j(k,p,\epsilon)$.
\end{proof}


\subsection{Main result on optimality}
\begin{lemma}\label{decreasing_V}
    Let $p, \hat{p} \in G$ 
such that $\hat{p} \geq p$. Suppose the goods are substitutes for the opponent. Then $\forall k \in \mathcal{M}, V(k, \hat{p}) \leq V(k,p)$.
\end{lemma}
\begin{proof}
  Let $k \in \mathcal{M}$ and $\kappa$ the constant strategy that consists in bidding $k$ at each round. Let $\mathcal{B}_k : f \mapsto (p \mapsto f(T(k,p,\epsilon)))$ and $V_k^{(0)} : p \mapsto w(k) - \langle k, p \rangle$.
  By virtue of \Cref{increasing_T}, if $f$ is non-increasing then $\mathcal{B}_k(f)$ is also non-increasing. 
    Thus for $ n \in \mathbb{Z}_{\geq 0}, \mathcal{B}^n_k(f) = \big(\mathcal{B}_k \circ \dots \circ \mathcal{B}_k)(f)$ is non-increasing. There exists $N \in \mathbb{Z}_{\geq 0}$ such that $V(k,p) = w(k) - \langle k, T^N(\kappa, p, \epsilon) \rangle = \mathcal{B}_k^N(V_k^{(0)})(k, p)$ (\Cref{termination}) and $\forall n \geq N, T^n(\kappa, p, \epsilon) = T^N(\kappa, p, \epsilon)$. Since we can bound the number of iterations of $\mathcal{B}_k$ regardless of the value of $p$ (e.g. by \Cref{max_price}), there exists $\Tilde{N} \in \mathbb{Z}_{\geq 0}$ such that $V(k,.) \equiv \mathcal{B}_{k}^{\Tilde{N}}(V_k^{(0)})$. Since $V_k^{(0)}$ is non-increasing, $p \mapsto V(k,p)$ is non-increasing on $\prod_{j = 1}^M (p_{\min, j} + \epsilon_j \mathbb{Z}_{\geq 0})$.
\end{proof}

\begin{theorem}\label{constant_optimality}
    Suppose the goods are substitutes for the opponent. For all $k \in \mathcal{M}$, for all prices $p \in G$, 
we have $\Tilde{V}(k,p) = W(k,p)$. 
\end{theorem}
\begin{proof}
    Let $k \in \mathcal{M}$ and $p \in \mathbb{R}^M_{\geq 0}\backslash\mathcal{T}$. First, if for all $j \in [M], k_j + \mathcal{D}_j(p) \leq m_j$, then $W(k,p) = w(k) - \langle k, p\rangle = \tilde{V}(k,p)$. Otherwise, note that 
    $W \geq V$. Hence, $W(k,p) = \max_{\|l\|_1 \leq \|k\|_1} W(l, T(k,p,\epsilon))$ is greater 
    or equal to $\Tilde{V}(k,p)=\max_{\|l\|_1 \leq \|k\|_1} V(l, T(k,p,\epsilon))$. To prove that $\tilde{V} \geq W$, we fix an optimal sequence of bids $(k^n)$ and the associated sequence of price $(p^n)$. 
    According to \Cref{termination}, there exists $N$ such that $\forall j \in [M], k^N_j + \mathcal{D}_j(p^N) \leq m_j$. We fix the smallest $N$ satisfying this condition, then $W(k,p) = W(k^0, p^0) = w(k^N) - \langle k^N, p^N \rangle = V(k^N, p^N)$. 
{If $N=0$, then $W(k,p) =V(k,p)\leq  \Tilde{V}(k, p)$. Otherwise, 
per \Cref{decreasing_V}, $V(k^N, p^N) \leq V(k^N, p^1)$. Therefore, $W(k,p) \leq V(k^N, p^1) \leq \Tilde{V}(k^0, p^0) = \Tilde{V}(k, p)$.} 
\end{proof}
\Cref{constant_optimality} ensures that, if the goods are ordinary substitutes for the opponent, the player has an optimal constant bid strategy.
Remark that goods are 
substitutes when $M=1$.

Furthermore, \Cref{constant_optimality} allows us to compute the value of the auction under perfect information. This amounts to computing 
{$k^* = \argmax_{k \in \mathcal{M}} V(k,p_{\min})$}.
The simulated values (see \Cref{fig:3Doutput}) suggest the two following questions:
\begin{figure}
    \centering
    \includegraphics[width=0.75\linewidth]{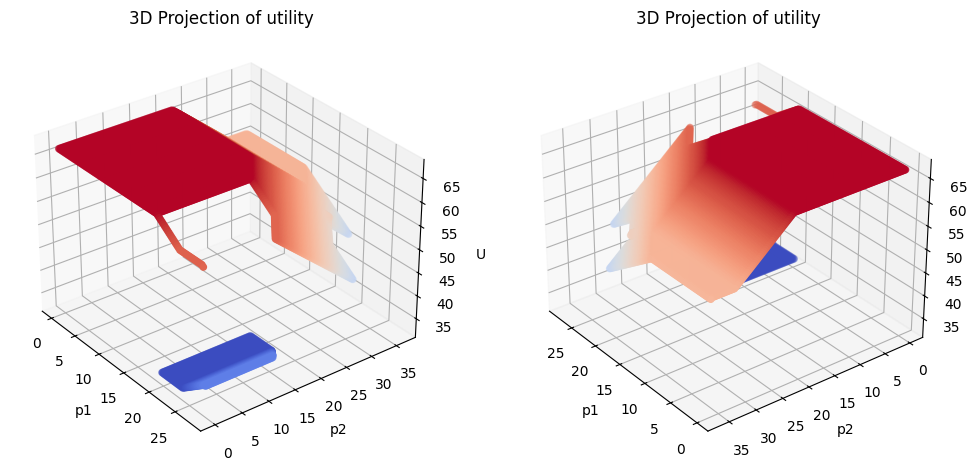}
        \vspace{-1.1em}
        \caption{3D visualization of $W$ for $M=2$}
        \vspace{-1.5em}
    \label{fig:3Doutput}
\end{figure}
\begin{enumerate}
    \item Is the value $W$ piecewise linear?
    \item Could the value $W$ have points of discontinuity?
\end{enumerate}
We show that the answer is positive in the next section.

\section{Continuous auction}\label{limit_auction}

To study the behavior of the value function $V(k,\cdot)$ under a constant bidding strategy, we now
consider the limit of the auction as
the price increment vector $\epsilon$ tends to zero, and establish the existence of a well-defined continuous-time auction.

Hereafter, we make the following assumption. 

\begin{assumption}\label{substitutes}
    The goods are substitutes for the opponent.
\end{assumption}


\subsection{Auction as a differential inclusion}

\begin{definition}
  Let $k \in \mathcal{M}$. Let
  \[ \gamma_k : p \in \mathbb{R}^M_{\geq 0}\backslash \mathcal{T} \mapsto \begin{pmatrix}
        \mathds{1}_{k_1 + \mathcal{D}_1(p) > m_1}\\
        \vdots\\
        \mathds{1}_{k_M + \mathcal{D}_M(p) > m_M}
  \end{pmatrix}\enspace
  \]
We define $\Gamma_k: p \in \mathbb{R}^M_{\geq 0}\mapsto \bigcap_{r > 0} \overline{\conv(\gamma_k(B(p, r)\backslash \mathcal{T}))}$, where $B(p,r)$ denotes the ball of radius $r$ centered at point $p$, and $\conv$ denotes the convex hull of a set.
\end{definition}

The drift $\gamma_k$ is used to define the continuous-time auction dynamics obtained 
by letting
$\epsilon \rightarrow 0$. Indeed, assume $\epsilon$ uniform {equal to $h$ and denote $p(nh)=p^n$. We then have $p(t+h) - p(t) = T(k,p(t),\epsilon) - p(t)= {h} \gamma_k(p(t))$ which} 
is a discrete approximation of $\gamma_k(p(t)) = \dot p(t)$. 

The map 
$\Gamma_k$ is known as the \textit{Filippov's regularization} of $\gamma_k$~(\cite{filippov_differential_1988, danca_class_2001}). It allows us to define the differential inclusion $\dot p \in \Gamma_k(p)$. When the demand is single-valued, the price dynamics is driven by $\gamma_k(p)$, whereas when the price lies on $\mathcal{T}$, it is driven by a convex combination of the adjacent $\gamma_k(p')$. This guarantees the existence of a solution.

\begin{theorem}\label{existence}
The differential inclusion
    \begin{equation}\label{inclusion}
    \tag{$I_{p}$}
        \dot \rho(s) \in \Gamma_k(\rho(s)),\qquad
        \rho(0) = p
    \end{equation}
    admits one absolutely continuous solution.
\end{theorem}
\begin{proof}
    Application of ~\cite[Theorem~5, \S 7, Chap.~2]{filippov_differential_1988} or ~\cite[Theorem~4, \S 2, Chap.~2]{aubin_differential_1984}.
\end{proof}

We next discuss the uniqueness of the solution $\rho$ which is needed to define a \textit{bona fide} continuous-time auction.

\subsection{Uniqueness of solution}\label{section_unique}

The uniqueness of the solution of a differential inclusion is known only in particular cases. Although $\gamma_k$ does not satisfy classical assumptions (such as Lipschitz continuity~\cite{filippov_differential_1988}, monotonicity \cite{aubin_differential_1984} or specific forms \cite{danca_class_2001}), we shall see the auction provides sufficient structure to derive uniqueness.
The following proposition shows that the discontinuity vector, i.e.~the jump of the piecewise-constant vector field when crossing the boundary of a {maximal} cell, belongs to the collection $\{ \pm e_k, e_i -e_j | k, i, j \in [M], i \neq j\}$. 
This is known as the unimodular system of $A_M$-type~\cite{danilov_discrete_2004}.
\begin{proposition}\label{shape_delta}
    (Prop. 3.10 of~\cite{baldwin_klemperer}) Let $\delta^+, \delta^- \in \mathcal{M}$ such that $\mathcal{D}^{-1}(\delta^+)\cap\mathcal{D}^{-1}(\delta^-) \neq \emptyset$ and $\delta^+ \neq \delta^-$. Then, $\exists (i,j) \in [M]^2, \delta^+ - \delta^- \in \{ e_i, e_i-e_j, -e_j \}$.
\end{proposition}

This forces the values of a solution's derivatives.

\begin{corollary}\label{u_delta}
    We denote by $u^\delta$ the unique value $\gamma_k(x)$ for any $x$ such that $\mathcal{D}(x) = \{\delta\}$.
    For $\delta^+, \delta^- \in \mathcal{M}$, we have:
    \begin{itemize}
    \item $\delta^+ - \delta^- = e_i - e_j$ $\implies u^{\delta^+} - u^{\delta^-} \in \{0, e_i, e_i-e_j, -e_j\}$;
    \item $\delta^+ - \delta^- = e_i \implies u^{\delta^+} - u^{\delta^-} \in \{0, e_i\}$;
    \item $\delta^+ - \delta^- = - e_j \implies u^{\delta^+} - u^{\delta^-} \in \{0, -e_j\}$.
    \end{itemize}
\end{corollary}
\begin{proof}
Immediate by the expression of $\gamma_k$.
\end{proof}

\Cref{shape_delta} also
constrains the neighboring demands.

\begin{corollary}\label{type_delta}
    Let $p \in \mathbb{R}^M_{\geq 0}$.  Then there exists $\delta_0\in \mathcal{M}$ such that one of the following statements is true:
    \begin{enumerate}
        \item[(i)] $\forall \delta \in \mathcal{D}(p), \big(\delta = \delta_0$ or $\exists i \in [M], \delta = \delta_0 - e_i\big)$.
        \item[(ii)] $\forall \delta \in \mathcal{D}(p), \big(\delta = \delta_0$ or $\exists i \in [M], \delta = \delta_0 + e_i\big)$.
    \end{enumerate}
\end{corollary}
\begin{proof}[Sketch of proof]
The result is proven by induction on the cardinal of $\mathcal{D}(p)$. The base case is straightforward. As for the induction step, we consider $\delta \in \mathcal{D}(p)$ and a price $p'$ such that $\mathcal{D}(p') = \mathcal{D}(p)\backslash\{\delta\}$. We then examine the compatibility of $\delta$ with the elements of $\mathcal{D}(p')$ whether it is of form (i) or (ii). \Cref{shape_delta} excludes certain forms of $\delta$, resulting in $\mathcal{D}(p)$ being of form (i) or (ii) for a well-chosen $\delta_0$.
\end{proof}

In the proof of the following theorem, we make use of a classical propriety of auctions: the law of aggregated demand.
\begin{lemma}\label{aggregated_demand}
(Law of aggregated demand (\cite[Prop.~3.E.4]{MWG}))
    Let $p, p' \in \mathbb{R}^M_{\geq 0}$. Let $\delta \in \mathcal{D}(p)$ and $\delta' \in \mathcal{D}(p')$. Then, $\langle p - p', \delta - \delta' \rangle \leq 0$.
\end{lemma}

\begin{theorem}\label{uniqueness}
    The differential inclusion \eqref{inclusion} admits exactly one absolutely continuous solution.
\end{theorem}
\begin{proof}
    Let $p \in \mathbb{R}^M_{\geq 0}$. Let $V$ be a neighborhood of $p$ such that $\mathcal{D}(V)\subseteq \mathcal{D}(p)$. Assume by contradiction that there exists $x$ and $y$ two distinct solutions of $(I_p)$ such that $x(t)\neq y(t)$ for $t\in (0,t_0)$. Without loss of generality, we assume that for all $t\leq t_0, x(t)$ and $y(t)$ remain in $V$.
    We consider the function $\mathcal{L}: t \in (0,t_0) \mapsto \frac{1}{2}\langle x(t) - y(t), x(t) - y(t) \rangle$. We shall show that it is non-increasing, i.e. show that $\frac{d}{dt}\mathcal{L}(t) = \langle \dot x(t) - \dot y(t), x(t) - y(t))\rangle \leq 0$.
    By definition of $\Gamma_k$, $\dot{x}(s)= \sum_{\delta\in \mathcal{D}(x(s))} \lambda^{\delta} u^{\delta}$ with $\lambda^\delta \geq 0$ with sum $1$ and similarly for $y$: $\dot{y}(s)= \sum_{\delta'\in \mathcal{D}(y(s))} \mu^{\delta'} u^{\delta'}$. Therefore, $\dot x (s) - \dot y(s) = \sum_{\substack{\delta \in \mathcal{D}(x(s)) \\ \delta' \in \mathcal{D}(y(s)) }}\lambda^\delta \mu^{\delta'}(u^\delta - u^{\delta'})$. Hence it suffices to prove that for all $\delta \in \mathcal{D}(x(s))$ and $\delta' \in \mathcal{D}(y(s))$, $\Delta(\delta,\delta') \coloneqq \langle u^\delta - u^{\delta'}, x(t) - y(t)\rangle$ is not positive. Let $\delta$ and $\delta'$ be such demands. 
    Note that if $u^\delta - u^{\delta'} = \delta - \delta'$, \Cref{aggregated_demand} gives $\Delta(\delta,\delta') \leq 0$. Also, if $u^\delta - u^{\delta'} = 0$ then $\Delta(\delta,\delta') = 0$. According to \Cref{u_delta}, the only case remaining is when $\delta - \delta' = e_i - e_j$ and $u^\delta - u^{\delta'} \in \{e_i, -e_j\}$. Per \Cref{type_delta}, suppose $\mathcal{D}(p)$ is of form {(ii) for some $\delta_0\in\mathcal{M}$, then} 
 $\delta = \delta_0 + e_i$ and $\delta' = \delta_0 + e_j$ (the proof is {similar} 
 if we assume $\mathcal{D}(p)$ of form {(i)}). Suppose $u^\delta - u^{\delta'} = e_i$. For $\delta'' \in \mathcal{D}(p)\backslash\{\delta,\delta'\}$, we write $\delta'' = \delta_0 + e_l$ with $l \in [M]\backslash\{i,j\}$ {or $\delta'' = \delta_0$}. Note that $\delta''_j = (\delta_0)_j = \delta_j$. Hence, $u^{\delta''}_j = u^{\delta}_j$. Furthermore, $u^{\delta'}_j = u^{\delta}_j$ since $u^\delta - u^{\delta'} = e_i$. Therefore, there exists a constant $u_j \in \{0, 1\}$ such that for all {$s\in (0,t_0)$ and $d \in \mathcal{D}(x(s))\cup\mathcal{D}(y(s))$}, $u^{d}_j = u_j$. By definition of $\Gamma_k$, this implies that {$\dot x_j(s) = \dot y_j(s)=u_j$, for $s\in (0,t_0)$}, thus granting $x_j(t) = y_j(t)$. Applying \Cref{aggregated_demand}, we have $x_i(t) - y_i(t) \leq x_j(t) - y_j(t) = 0$ therefore $\Delta(\delta, \delta') \leq 0$.
{The case $u^\delta - u^{\delta'} =- e_j$ is symmetrical.}
    All in all, the {Euclidian} norm 
    of $x(t) - y(t)$ is nonincreasing on $(0,t_0)$ and has value $0$ at $t=0$ therefore for all $t\in [0, t_0),$ $x(t) = y(t)$.
\end{proof}
Theorems~\ref{existence} and~\ref{uniqueness} guarantee the existence and uniqueness of the solution to ($I_{p_{\min}}$). The continuous-time auction is thus well-defined, in the sense that each initial price determines a unique price trajectory. 
Furthermore, this trajectory depends continuously on $p_{\min}$~\cite[Corollary~1 and~2, \S 8, Chap.~2]{filippov_differential_1988}. 

\subsection{Value in continuous auction}

Since the auction price trajectory is unique, we extend the definition of the values $V$ and $\Tilde{V}$ introduced in \Cref{def_v}:

\begin{definition}
    For $k \in \mathcal{M}$ and $p \in \mathbb{R}^M_{\geq 0}$, we set $\mathcal{V}(k, p) = w(k) - \langle k, P_k(p) \rangle$ where $P_k(p)$ is the final price of the auction (i.e.\ the limit when $t \rightarrow \infty$ of the unique solution of the \eqref{inclusion}). 
    
    We also define $\Tilde{\mathcal{V}}(k,p) \coloneqq  w(k) - \langle k, p \rangle \text{ if }\forall j \in [M], k_j + \mathcal{D}_j(p) \leq m_j\text{ and }\max_{\substack{l \in \mathcal{M},\, \|l\|_1 \leq \|k\|_1}}\mathcal{V}(l,p)$ otherwise.
\end{definition}

The following lemma shows that every cell at the interface of a maximal cell
is crossed at most once. Hence, every maximal cell is crossed a finite number of times.

 \begin{lemma}\label{cross}
   Let $k \in \mathcal{M}$ and $p \in \mathbb{R}^M_{\geq 0}$. Let $\rho$ be the solution of \eqref{inclusion} on $[0, T]$. 
   We define $\Trav(\delta^-, \delta^+)$ as the set of crossing times between the maximal cells where $\delta^+$ and $\delta^-$ are the only demands on their interiors. We denote by $u^+ \coloneqq  \gamma_k(\mathcal{D}^{-1}(\delta^+)\backslash\mathcal{T})$ and $u^- \coloneqq  \gamma_k(\mathcal{D}^{-1}(\delta^-)\backslash\mathcal{T})$.
   If $u^+ \neq u^-$, then $\#\Trav(\delta^-, \delta^+) \leq 1$.
 \end{lemma}
 \begin{proof}
    Suppose $B\coloneqq\mathcal{D}^{-1}(\delta^-)\cap\mathcal{D}^{-1}(\delta^+) \neq \emptyset$ otherwise the result is immediate. If $t \in \Trav(\delta^-, \delta^+)$ then $\rho(t)\in B$ and there exists $h > 0$ such that $\forall s \in [t-h, t)$, $\rho(s) \in \mathcal{D}^{-1}(\delta^-)\backslash\mathcal{D}^{-1}(\delta^+)$ and $\forall s \in (t, t+h]$, $ \rho(s) \in \mathcal{D}^{-1}(\delta^+)\backslash\mathcal{D}^{-1}(\delta^-)$. 
    By contradiction, assume $\#\Trav(\delta^-, \delta^+) > 1$. Let $(t_1, t_2) \in \Trav(\delta^-, \delta^+)^2$ with $t_2 > t_1$. We fix the associated $h_1 > 0$ and $h_2 > 0$. Let $h = \min(h_1, h_2) > 0$. Let $i\neq j$ such that $\delta^- - \delta^+ \in \{ e_i, e_i - e_j\}$ (\Cref{shape_delta}). 
    Suppose $\delta^- - \delta^+ = e_i$ then $B \subseteq \{x \in \mathbb{R}^M : x_i = v(\delta^-) - v(\delta^+) \}$. Thus $\langle \rho(t_2 -h) - \rho(t_1), e_i \rangle =$ $ \rho_i(t_2 -h) - \rho_i(t_1) < (v(\delta^-) - v(\delta^+)) - (v(\delta^-) - v(\delta^+)) = 0$. However, $\dot \rho \geq 0$ therefore $\rho(t_2 - h) \geq \rho(t_1)$ which is a contradiction.
    Suppose $\delta^- - \delta^+ = e_i - e_j$ then $B \subseteq \{x \in \mathbb{R}^M : x_i - x_j = v(\delta^-) - v(\delta^+) \}$. Note that $\rho_i(t_2 -h) - \rho_j(t_2 - h) < v(\delta^-) - v(\delta^+)$ and $\rho_i(t_2) - \rho_j(t_2) = v(\delta^-) - v(\delta^+)$. Since $u^- \in \{0, 1\}^M$, we necessarily have $u^-_i = 1$ and $u^-_j = 0$. Symmetrically, since $\rho_i(t_1+h) - \rho_j(t_1+h) > v(\delta^-) - v(\delta^+)$, necessarily $u^+_i = 1$ and $u^+_j = 0$. $\forall l \notin \{i, j \}, u^+_l = u^-_l$ because $\delta^- - \delta^+ = e_i - e_j$. Therefore, $u^- = u^+$ which is a contradiction. 
 \end{proof}

\begin{definition}
    Let $k \in \mathcal{M}$. Let $\Sigma$ be the set of cells defined by $\mathcal{T}$ (not necessarily maximal), and $\Sigma^* \coloneqq  \bigcup_{n>0}\Sigma^n$. For every finite sequence $\sigma \in \Sigma^*$, we define $\mathcal{P}_{\sigma}^{k}$ as the set of initial prices $p$ for which the sequence of minimal cells visited by the solution of \eqref{inclusion} is exactly $\sigma$.
\end{definition}

Note that 
$\rho$ crosses or slides through a finite number of cells (\Cref{cross}). Therefore, the set $\{\sigma \in \Sigma^* : \mathcal{P}^k_\sigma \neq \emptyset \}$ is finite. Moreover $\bigcup_{\sigma \in \Sigma^*} \mathcal{P}_{\sigma}^{k} = \mathbb{R}^M_{\geq 0}$ since it forms a partition of the set of prices. Those two observations are crucial for the proof of the piecewise linear behavior of $\tilde{\mathcal{V}}$.

We next make use of the notion of {\em semilinear set}. We call {\em polyhedra} a set defined by finitely many weak or strict affine inequalities.

\begin{definition}[See e.g.~\cite{van_den_dries_tame_1998}]
    A set is {\em semilinear} if it is a finite union of polyhedra.
A function is {\em semilinear} if its graph is semilinear. 
\end{definition}
Thus, the semilinear functions are exactly the piecewise linear functions (not necessarily continuous).
The family of semilinear sets is closed  under 
union, taking the complement, and projection. They constitute
the simplest example of o-minimal structure.
We refer the reader to the monograph of Van den Dries~\cite{van_den_dries_tame_1998} for background.
These tools will allow us to prove the semilinearity of $\mathcal{V}$ and $\tilde{\mathcal{V}}$.

\begin{lemma}\label{lem-seminice}
    Let $k \in \mathcal{P}$. For all $\sigma \in \Sigma^*, \mathcal{P}_{\sigma}^{k}$ is semi-linear.
\end{lemma}
\begin{proof}
  We prove it by induction on the length of the sequence $\sigma$. Suppose that $\sigma \in \Sigma$,
  i.e. $\sigma$ is a single cell.
  Let $p \in \mathcal{P}_{\sigma}^{k}$. The solution of \eqref{inclusion} passes through exactly one cell. If this
  cell has maximal dimension, the solution navigates in the interior of this cell, and its speed $\dot \rho$
  is equal to a constant vector $c$ by definition. If the dimension of this cell is non-maximal, it follows from the uniqueness of the trajectory (proof of \Cref{uniqueness}) that $\dot \rho$ is still equal to some constant vector $c$, sliding along the cell. 
Hence, we see that
$\mathcal{P}_{\sigma}^{k}=\{p \in \sigma\mid p+\mathbb{R}_{\geq 0}c \in \sigma\}$. This entails
that $c$ is zero (because prices cannot increase indefinitely),
and then $\mathcal{P}_{\sigma}^{k}=\sigma$ is semilinear.
Let $n > 0$ and assume that for all $\sigma \in \Sigma^n, \mathcal{P}_{\sigma}^{k}$ is semi-linear. Let $\sigma \in \Sigma^{n+1}$. Note that $\sigma = (\xi,\sigma')$ with $\sigma' \in\Sigma^n$. Therefore, for $p \in \mathcal{P}_{\sigma}^{k}$, a solution encounters $\xi$ then falls in $\mathcal{P}_{\sigma'}^{k}$. 
Let $c$ be the constant derivative $\dot{\rho}$ in the cell $\xi$. If $c$ is transverse to $\xi$, then
the trajectory $\rho$ crosses instantaneously $\xi$, so that
$\mathcal{P}_{\sigma}^k= \{p\in\xi \mid p + t c\in \mathcal{P}_{\sigma'}^k \text{ for }t>0\text{ small enough}\}$. This set is semilinear~\cite[Coro.~7.8.]{van_den_dries_tame_1998}.
If $c$ slides along $\xi$, the trajectory $\rho$ stays in the cell $\xi$ until it meets the next cell.
Then, $\mathcal{P}_{\sigma}^{k} = \{p\mid \exists t_1>0, p+[0,t_1)c \in \xi, p+t_1 c\in \mathcal{P}_{\sigma'}^{k}\}$, and it follows that $\mathcal{P}_{\sigma}^{k}$ is semi-linear ({\em ibid.}~\cite[Coro.~7.8.]{van_den_dries_tame_1998}).
\end{proof}
\begin{lemma}\label{lem-onsemi}
  Let $k \in \mathcal{M}$. For all $\sigma \in \Sigma^*$, the function $p \in \mathcal{P}_{\sigma}^{k} \mapsto \mathcal{V}(k,p)$ is semilinear.
\end{lemma}
\begin{proof}[Sketch of proof]
    Arguing as in the proof of \Cref{lem-seminice}, we prove that the function $f$ that maps $p \in \mathcal{P}_{\sigma}^k$ to $\rho(\tau)$, the value of the solution of \eqref{inclusion} at the end of the auction is semilinear. It follows that $\mathcal{V}(k,p) = w(k) - \langle k, f(p)\rangle$ is a semi-linear function of $p\in \mathcal{P}_\sigma^k$.
\end{proof}

\begin{proposition}
    The function $p \in \mathbb{R}^M_{\geq 0} \mapsto {\mathcal{V}}(k,p)$ is semi-linear for all $k \in \mathcal{M}$.
\end{proposition}
\begin{proof}
  This follows from~\Cref{lem-onsemi}
  and from the fact the partition $\bigcup_{\sigma \in \Sigma^*} \mathcal{P}_{\sigma}^{k} = \mathbb{R}^M_{\geq 0}$
  is finite.
\end{proof}

\begin{theorem}\label{th-valueissemilinear}
    The function $p \in \mathbb{R}^M_{\geq 0} \mapsto \Tilde{\mathcal{V}}(k,p)$ is semi-linear for all $k \in \mathcal{M}$.
\end{theorem}
\begin{proof}
    Let $k \in \mathcal{M}$. For all $l \in \mathcal{M}$ such that $ \|l\|_1 \leq \|k\|_1$, $\mathcal{V}(l,.)$ is semi-linear. Hence, as the maximum of semi-linear functions, $\Tilde{\mathcal{V}}(k,.)$ is also semi-linear.
\end{proof}

We can thus conclude that the value over constant bidding strategies is piecewise linear. Observe that the discontinuity emerges from taking the maximum over admissible strategies in $\Tilde{\mathcal{V}}$, a gap can thus happen when the maximum changes between two close prices.

\begin{corollary}\label{coro-approx}
  The value function $W_\epsilon$ of the auction 
  with a price increment $\epsilon$ converges as $\epsilon$ tends to zero to the value function $\tilde{\mathcal{V}}$ of the continuous-time auction.
  \end{corollary}
\begin{proof}
  This follows Euler methods for differential inclusions~\cite[Theorem~3.1.7]{Clarke1990} (see also~\cite[Theorem~2.2]{Dontchev1992}). 
  \end{proof}
\section{Application to the multiband residual 2017 Australian auction}


We model a simplified version of the 2017 Australian multiband residual lot auction~\cite{auction_guide}. The Australian Communications and Media Authority organized an SCA for blocks belonging to the 1800 MHz, 2 GHz, 2.3 GHz and 3.4 GHz bandwidths. 
They were arranged into geographical lots on which the competitors could bid. We focus particularly on Stage 1 of the auction where 
$5$ lots of blocks belonging to the $1800$ MHz band and $3$ lots of the $2$ GHz bands were at stake~\cite{ radiocommunications_australia}.
We gather the result of this phase in \Cref{fig:results} according to the available data~\cite{auction_results}. 

\begin{figure}[!htbp]
\centering
    \tiny{
\begin{tabular}{ |c|c|c|c|c| } 
\hline
Bands & Geographic Area & Starting Price (AUD) & Ending Price (AUD) & Winning Bid \\
\hline
\multirow{5}{4em}{$1800$ MHz} & Dubbo & $252,000$& $252,000$& PDP/TPG  \\ 
& \cellcolor{lightgray} Mackay & \cellcolor{lightgray} $1,120,000$ & \cellcolor{lightgray}$1,360,000$ & \cellcolor{lightgray} TPG  \\ 
& Maryborough & $4,302,000$& $4,302,000$& Telstra \\ 
& WA & $235,000$& $235,000$& Vodafone \\ 
& \cellcolor{lightgray} Tasmania & \cellcolor{lightgray}$410,000$ & \cellcolor{lightgray}$970,000$ & \cellcolor{lightgray}TPG  \\ 
\hline
\multirow{3}{4em}{$2$ GHz} & Canberra & $2,147,000$& $2,147,000$& Telstra \\
 & \cellcolor{lightgray} Darwin & \cellcolor{lightgray}$630,000$& \cellcolor{lightgray}$2,451,000$& \cellcolor{lightgray} Vodafone \\ 
& \cellcolor{lightgray} Hobart & \cellcolor{lightgray}$1,181,000$& \cellcolor{lightgray}$4,551,000$& \cellcolor{lightgray}Vodafone \\ 
\hline
\end{tabular}}
    \caption{Results of the multiband residual auction in Australia (2017). The studied lots are highlighted in light gray.}        \vspace{-1.5em}
    \label{fig:results}
\end{figure}

We observe that the lots Dubbo, Maryborough, WA and Canberra were sold at their starting prices. This shows that the five operators (NBN Co, Optus, Telstra, TPG and Vodafone) were not competing on those blocks. We thus focus on the Mackay and Tasmania lots in the $1800$ MHz band and the Darwin and Hobart lots in the $2$ GHz. Note that these bands were not aggregated since they have different starting prices. Nonetheless, we will consider the two bandwidths as two categories. We could model the auction with four different categories, however in order to plot meaningful illustrations, we chose to simplify the model to a 2-dimensional auction. 

We define an auction with $M = 2$ and $m_1 = m_2 = 2$. The initial price will be taken as an average of the initial prices of the category: $p_{\min} = (765, 905.5)$. Let $p_{\max} = (1165, 3501)$ be the final observed price of the auction. We notice that the winner of the $2$ GHz lots was Vodafone (resp.\ TPG wins the $1800$ MHz lots). If the allocation were optimal, according to \Cref{constant_optimality}, the constant strategy $(0, 2)$ (resp.\ $(2,0)$) should have been optimal for Vodafone (resp.\ TPG). Hence, at the start of the auction, the prices would not have been raised as $\mathcal{D}_{\text{TPG}}(p_{\min}) + \mathcal{D}_{\text{Vodafone}}(p_{\min}) \leq m$.

\begin{figure}
    \centering
    \includegraphics[width=0.95\linewidth]{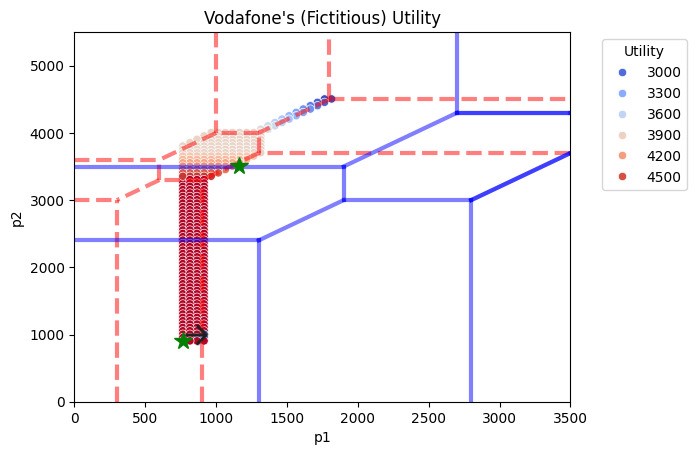}
    \vspace{-1.5em}
    \caption{Value of $W_{\text{Vodafone}}$ at each reachable price assuming TPG is SB.}
    \vspace{-1.1em}
    \label{fig:vodafone}
\end{figure}

\begin{figure}
    \centering
    \includegraphics[width=0.95\linewidth]{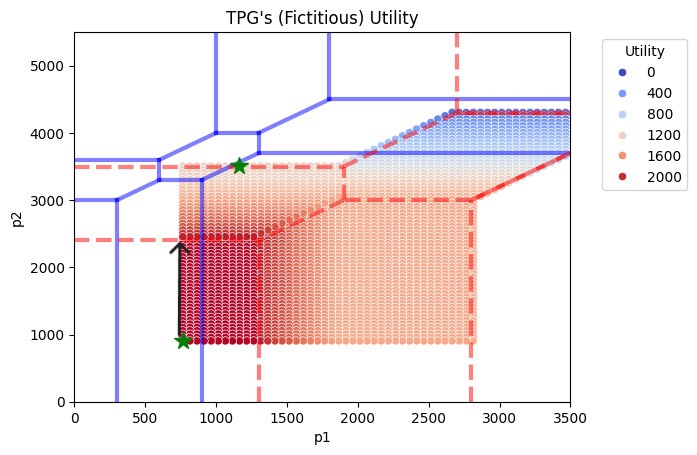}
    \vspace{-1.5em}
    \caption{Value of $W_{\text{TPG}}$ at each reachable price assuming Vodafone is SB.}
    \vspace{-1.5em}
    \label{fig:TPG}
\end{figure}

In order to model what the auction would have looked like with perfect information, we draw random valuations for TPG and Vodafone such that $\max_{k \in \mathcal{M}} (v_{\text{Vodafone}}(k) - \langle k, p_{\max} \rangle)$ $= v_{\text{Vodafone}}(0, 2) - \langle (0, 2), p_{\max}\rangle$ (resp.\ for TPG with $(0,2)$) and such that the price $p_{\max}$ is reachable. We display the results in \Cref{fig:vodafone} and \Cref{fig:TPG}. These graphs illustrate the projection of the maximal payoff (using constant strategy) of a perfect informed player against a SB opponent. Each reachable price of the auction is represented by a dot, colored according to the value at this point. 
We show the tropical hypersurfaces of both competitors, the one of the opponent being dashed. The two stars represent the prices $p_{\min}$ and $p_{\max}$.

Note that the reachable prices depend on what competitor is SB. Indeed, the dynamics of prices as presented in the previous section depends on the opponent's tropical hypersurface. Therefore, a different competitor would lead to a different dynamics.


We observe that if TPG was SB then Vodafone could have maximize their payoff by bidding $(2, 0)$. In this case, the dynamics of prices would be immediate to plot on \Cref{fig:vodafone}: we represent it by an arrow. Indeed, this bid would increase the price to $\sim(1000, 1000)$ and result in a maximal payoff ($\sim4500$ in the displayed example \Cref{fig:vodafone}). Rising the price to $p_{\max}$ decreases this payoff to $\sim3900$.

On the other hand, if Vodafone was SB, then TPG could have secured a maximal payoff of $\sim2000$ by bidding $(0, 1)$ (see the arrow in \Cref{fig:TPG} for the dynamics). This is worth-mentioning since, despite $(0,2)$ being the optimal demand at price $p_{\max}$, reducing one's demand yields a higher payoff by decreasing the price. We also note that the payoff at $p_{\max}$ would have been almost half the optimal value.

It is worth noticing is that both scenarios are compatible in the sense that if each player has perfect information and believes their opponent to play SB, then they would respectively bid on $(2,0)$ and $(0, 1)$ ending the auction instantly. 
Although this allocation is optimal for a player in a perfect information setting against an SB opponent, it does not necessary lead to an equilibrium. In this example, bidding $(0, 2)$ for TPG when Vodafone bids $(2,0)$ dominates bidding $(0,1)$. This allocation also highlights one drawback of the SCA which is that items can remain unsold.

\section{Conclusion}

We studied a multidimensional SCA in perfect information and detailed its structure under the ordinary substitutes condition. Our results are presented for a 2-player 
auction. They carry over to the $N$-player case since
Straightforward bidders can be aggregated~\cite{zeroual_2025}. 
In future work, we plan to leverage the present polyhedral approach
to tackle the problem in imperfect information. 

Our study 
completes the understanding of the ordinary substitutes condition which was used to ensure the existence of equilibria~(\cite{milgrom_substitute_2009, baldwin_klemperer}) and is essential to ensure optimality of constant strategies. In particular, 
we deduced that the value of the player is piecewise-linear in the continuous-time auction.
This reveals an intrinsic behavior of the auction,
{\em independent} of the price increment. 
Hence, we can device an algorithm to compute the optimal strategy that does not depend of the price increment thanks to the continuous-time auction model.
This
is relevant to the telecommunication industry since operators usually have an estimate of their competitors' valuations and wish to grasp the characteristics of optimal strategies as well as to compute such strategies.
We show how this could been used by 
a retrospective analysis of
the 2017 multiband residual Australian auction. 

Future work will investigate the extension of this approach to other formats
such as the Simultaneous Multi-Round Auction (SMRA) or the Clock-SMRA Hybrid.








\bibliographystyle{IEEEtran}
\bibliography{ECC}

\end{document}